\documentclass[useAMS,usenatbib]{mn2e}
\input psfig.sty

\def \aj {AJ}
\def \apj {ApJ}
\def \apjl {ApJL}
\def \mnras {MNRAS}
\def \apjs {ApJS}

\def \etal {et~al.~}

\def \spose#1{\hbox  to 0pt{#1\hss}}  
\def \lta{\mathrel{\spose{\lower 3pt\hbox{$\sim$}}\raise  2.0pt\hbox{$<$}}}
\def \gta{\mathrel{\spose{\lower  3pt\hbox{$\sim$}}\raise 2.0pt\hbox{$>$}}}

\def \kmsmpc {\>{\rm km}\,{\rm s}^{-1}\,{\rm Mpc}^{-1}}
\def \kms {\ifmmode  \,\rm km\,s^{-1} \else $\,\rm km\,s^{-1}  $ \fi }
\def \kpc {\ifmmode  {\rm kpc}  \else ${\rm  kpc}$ \fi  }  
\def \pc {\ifmmode  {\rm pc}  \else ${\rm  pc}$ \fi  }  
\def \Msun {\ifmmode M_{\odot} \else $M_{\odot}$ \fi} 
\def \Mstar {\ifmmode M_{\rm star} \else $M_{\rm star}$ \fi} 
\def \Msps {\ifmmode M_{\rm SPS} \else $M_{\rm SPS}$ \fi} 
\def \Mdyn {\ifmmode M_{\rm dyn} \else $M_{\rm dyn}$ \fi} 

\def \DeltaIMF {\ifmmode \Delta_{\rm IMF} \else $\Delta_{\rm IMF}$ \fi}
\def \dvr {\ifmmode \partial_{\rm VR} \else $\partial_{\rm VR}$ \fi} 
\def \sigmaap {\ifmmode \sigma_{\rm ap} \else $\sigma_{\rm ap}$ \fi} 
\def \Re {\ifmmode R_{\rm e} \else $R_{\rm e}$ \fi} 

\title[The IMF of dense early-type galaxies] {Evidence for a non-universal stellar initial mass function in low-redshift high-density early-type galaxies}

\author[Dutton et al.]  {Aaron  A.
  Dutton$^{1,2,3}$\thanks{dutton@mpia.de}\thanks{CITA National Fellow}, J. Trevor Mendel$^1$, Luc Simard$^4$\\  
  $^1$Dept. of Physics and Astronomy, University of Victoria, Victoria, B.C., V8P 5C2, Canada\\
  $^2$Dept. of Physics, University of California, Santa Barbara, CA 93106, USA\\
  $^3$Max Planck Institute for Astronomy, K\"onigstuhl 17, 69117, Heidelberg, Germany\\
  $^4$Herzberg Institute of Astrophysics, National Research Council of Canada, 5071 West Saanich road, Victoria, B.C., V9E 2E7, Canada\\}
\begin{document}
             
\date{Accepted 2012 January 27. Received 2012 January 24; in original form 2011 November 12}
             
\pagerange{\pageref{firstpage}--\pageref{lastpage}}\pubyear{2012}

\maketitle           

\label{firstpage}
             
%%%%%%%%%%%%%%%%%%%%%%%%%%%%%%%%%%%%%%%%%%%%%%%%%%%%%%%%%%%%%%%%%%%%%%

\begin{abstract}
  We determine an absolute calibration of stellar mass-to-light ratios
  for the densest $\simeq 3\%$ of early-type galaxies in the local
  universe (redshift $z\simeq 0.08$) from SDSS DR7.  This sample of
  $\sim 4000$ galaxies has, assuming a Chabrier IMF, effective stellar
  surface densities, $\Sigma_{\rm e} > 2500 \Msun/\pc^2$, stellar
  population synthesis (SPS) stellar masses
  $\log_{10}(\Msps/\Msun)<10.8$, and aperture velocity dispersions of
  $\sigmaap=168^{+37}_{-34}\kms$ (68\% range).
  In contrast to typical early-type galaxies, we show that these dense
  early-type galaxies follow the virial fundamental plane, which
  suggests that mass-follows-light. With the additional assumption that
  any dark matter does not follow the light, the dynamical masses of
  dense galaxies provide a direct measurement of stellar masses. 
  Our dynamical masses ($\Mdyn$), obtained from the spherical Jeans
  equations, are only weakly sensitive to the choice of anisotropy
  ($\beta$) due to the relatively large aperture of the SDSS fiber for
  these galaxies: $R_{\rm ap} \simeq 1.5 \Re$.
  Assuming isotropic orbits ($\beta=0$) we find a median
  $\log_{10}(\Mdyn/\Msps)=0.233\pm0.003$, consistent with a Salpeter
  IMF, while more bottom heavy IMFs and standard Milky-Way IMFs are
  strongly disfavored.  Our results are consistent with, but do not
  require, a dependence of the IMF on dynamical mass or velocity
  dispersion.  We find evidence for a color dependence to the IMF such
  that redder galaxies have heavier IMFs with $\Mdyn/\Msps\propto
  (g-r)^{1.13\pm0.09}$. This may reflect a more fundamental dependence
  of the IMF on the age or metallicity of a stellar population, or the
  density at which the stars formed.
\end{abstract}

\begin{keywords}
  galaxies: elliptical and lenticular, cD -- galaxies: fundamental
  parameters -- galaxies: kinematics and dynamics -- galaxies:
  structure
\end{keywords}

\setcounter{footnote}{1}

%%%%%%%%%%%%%%%%%%%%%%%%%%%%%%%%%%%%%%%%%%%%%%%%%%%%%%%%%%%%%%%%%%%%%%
%% SECTION 1: INTRODUCTION
%%%%%%%%%%%%%%%%%%%%%%%%%%%%%%%%%%%%%%%%%%%%%%%%%%%%%%%%%%%%%%%%%%%%%%
\section{Introduction}
\label{sec:intro}

The stellar initial mass function (IMF) is a fundamental property of a
stellar population, with wide ranging implications for many areas of
astrophysics.  Observations in the Galactic disk suggest that the IMF
has a power-law shape, $dN/dm \propto m^{-x}$, with $x\simeq -2.35$ at
masses above $m\simeq 1 \Msun$ (Salpeter 1955), and that it turns over
at lower masses (Kroupa 2001; Chabrier 2003).  For simplicity, the IMF
is generally assumed to be universal, although there is increasing
evidence that the IMF depends on the mass (or velocity) of a galaxy.
Spiral galaxies require IMFs lighter than Salpeter (Bell \& de Jong
2001; Dutton \etal 2011 a,b; Brewer \etal 2012) and maybe even lighter
than Chabrier (Bershady \etal 2011). While elliptical galaxies favor
Salpeter-type IMFs (Auger \etal 2010), and possibly bottom heavy
($x\simeq -3$) IMFs in the most massive ellipticals (van Dokkum \&
Conroy 2010).

Total mass measurements from dynamics or strong lensing only give an
upper limit to stellar mass-to-light ratios due to the unknown
dark matter fraction.  However, if we can identify a population of
galaxies that are baryon dominated (i.e., the dark matter fraction
within a specific radius $< 10\%$), these can be used to provide a
direct dynamical constraint to stellar mass-to-light ratios, and hence
the IMF.

In this Letter we show that early-type galaxies with the highest
stellar surface densities within an effective radius have a
Fundamental Plane (FP; Dressler \etal 1987; Djorgovski \& Davis 1987)
correlation that is consistent with the virial prediction: At fixed
SPS stellar mass, velocity dispersions and half-light radii scale as
$\sigma_{\rm ap} \propto R_e^{-1/2}$. This is strong evidence that the
dynamical masses are proportional to the stellar masses.  Under the
additional assumption that any dark matter does not trace the light,
the dynamical masses are equal to the stellar masses.  Here we use
spherical Jeans models to calculate dynamical masses and compare these
with SPS stellar masses.
We adopt a cosmology with $\Omega_{\Lambda}=0.7$, $\Omega_{\rm m}=0.3$
and $H_0=70 \kmsmpc$.

%%%%%%%%%%%%%%%%%%%%%%%%%%%%%%%%%%%%%%%%%%%%%%%%%%%%%%%%%%%%%%%%%%%%%%%%%%%%
%% SECTION 2 SAMPLE SELECTION
%%%%%%%%%%%%%%%%%%%%%%%%%%%%%%%%%%%%%%%%%%%%%%%%%%%%%%%%%%%%%%%%%%%%%%%%%%%%%%%

\vspace{-0.6cm}
\section{SAMPLE SELECTION}
\label{sec:sample}
Our selection of early-type galaxies is designed to select non-star
forming galaxies regardless of their morphology or structure, and is
the same as in Dutton \etal (2011a).  Our starting sample consists of
$\sim 685 000$ galaxies from the Sloan Digital Sky Survey (SDSS; York
\etal 2000) data release seven (DR7) (Abazajian \etal 2009) with
spectroscopic redshifts ($0.005 \le z \le 0.2$), stellar velocity
dispersions, $\sigma_{\rm ap}$, from the ``Princeton'' catalog,
structural parameters (using the S\'ersic $n=4$ plus $n=1$ fits) from
Simard \etal (2011), and stellar masses computed assuming a Chabrier
(2003) IMF from the MPA/JHU group (available at
http://www.mpa-garching.mpg.de/SDSS/DR7/).  We select galaxies that
have been spectroscopically classified as early-type (eCLASS $< 0$),
have red $(g-r)$ colors, and have minor-to-major axis ratios greater
than 0.5.  We also apply redshift dependent minimum stellar mass.  Our
selection yields $\sim 140 000$ early-type galaxies at redshifts of
$z\simeq 0.079^{+0.037}_{-0.030}$ (68\% range).

Using the stellar masses, $M_{\rm SPS}$, and the circularized $r-$band
effective radii, $R_{\rm e}$, we compute the average surface density
within the effective radius: $\Sigma_{\rm e}= M_{\rm SPS}/(2\pi R_{\rm
  e}^2)$.  We select a sample of dense galaxies with the criteria
$\Sigma_{\rm e} > 2500 \Msun \pc^2$.  This yields $\sim 5000$
galaxies, i.e., $\simeq 3.5\%$ of our full early-type galaxy sample.
While the exact value of this minimum density is somewhat arbitrary,
its approximate value will become apparent below.
In order to minimize the impact of galaxies scattered to high stellar
densities by the $\simeq 0.1$ dex measurement errors on stellar
masses, we clean the sample by using the relation between $\Msps/L_r$
and $\sigma_{\rm ap}$. This relation has a slope of $\simeq 0.4$ and a
scatter of $\simeq 0.06$ dex.  We remove the outlier galaxies that are
more than $\pm3\sigma$ offset from the median relation. This removes
$\simeq 400$ galaxies leaving a dense sample of $\simeq 4600$.

\begin{figure}
\centerline{
\psfig{figure=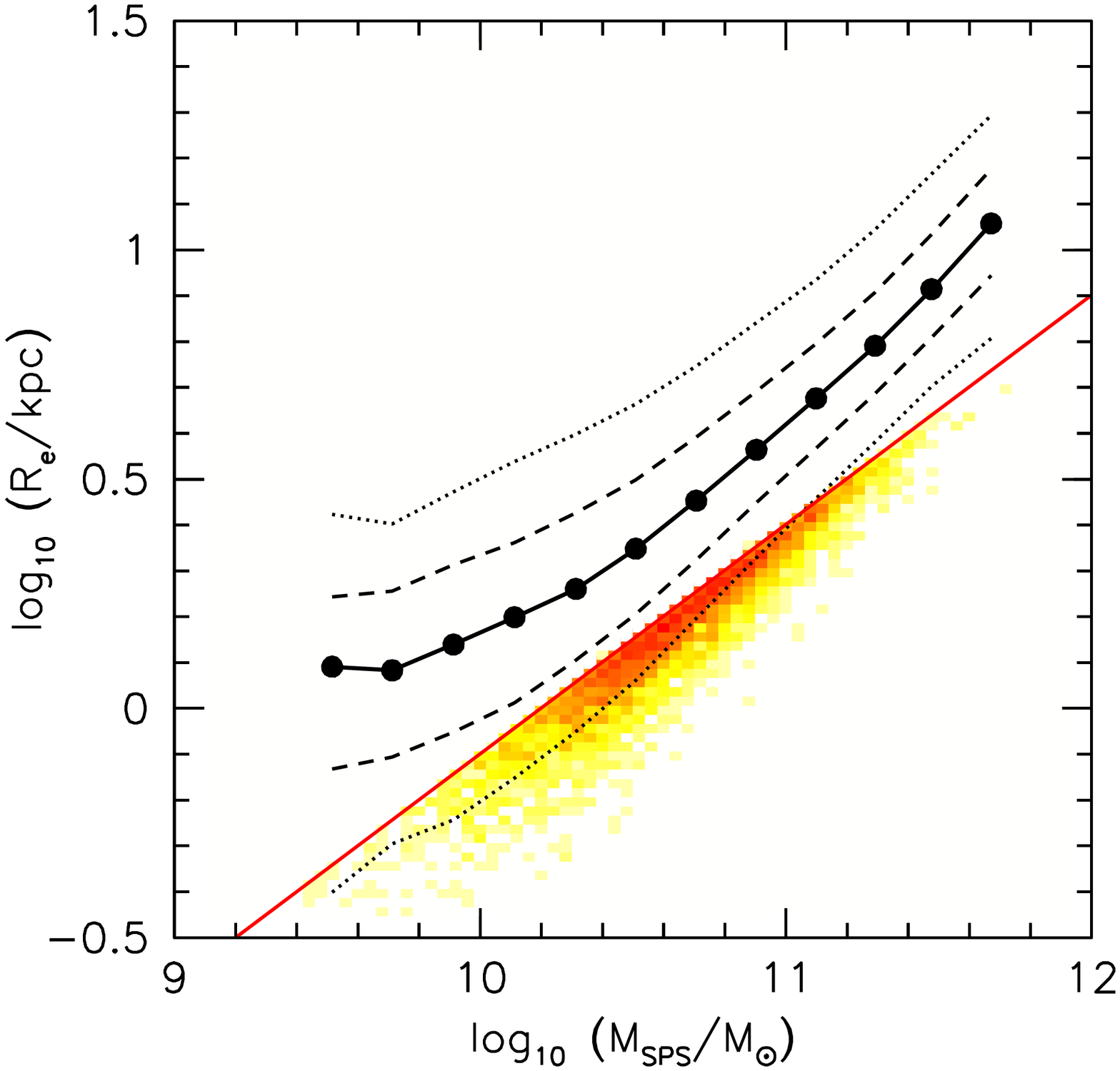,width=0.35\textwidth}
}
\centerline{
\psfig{figure=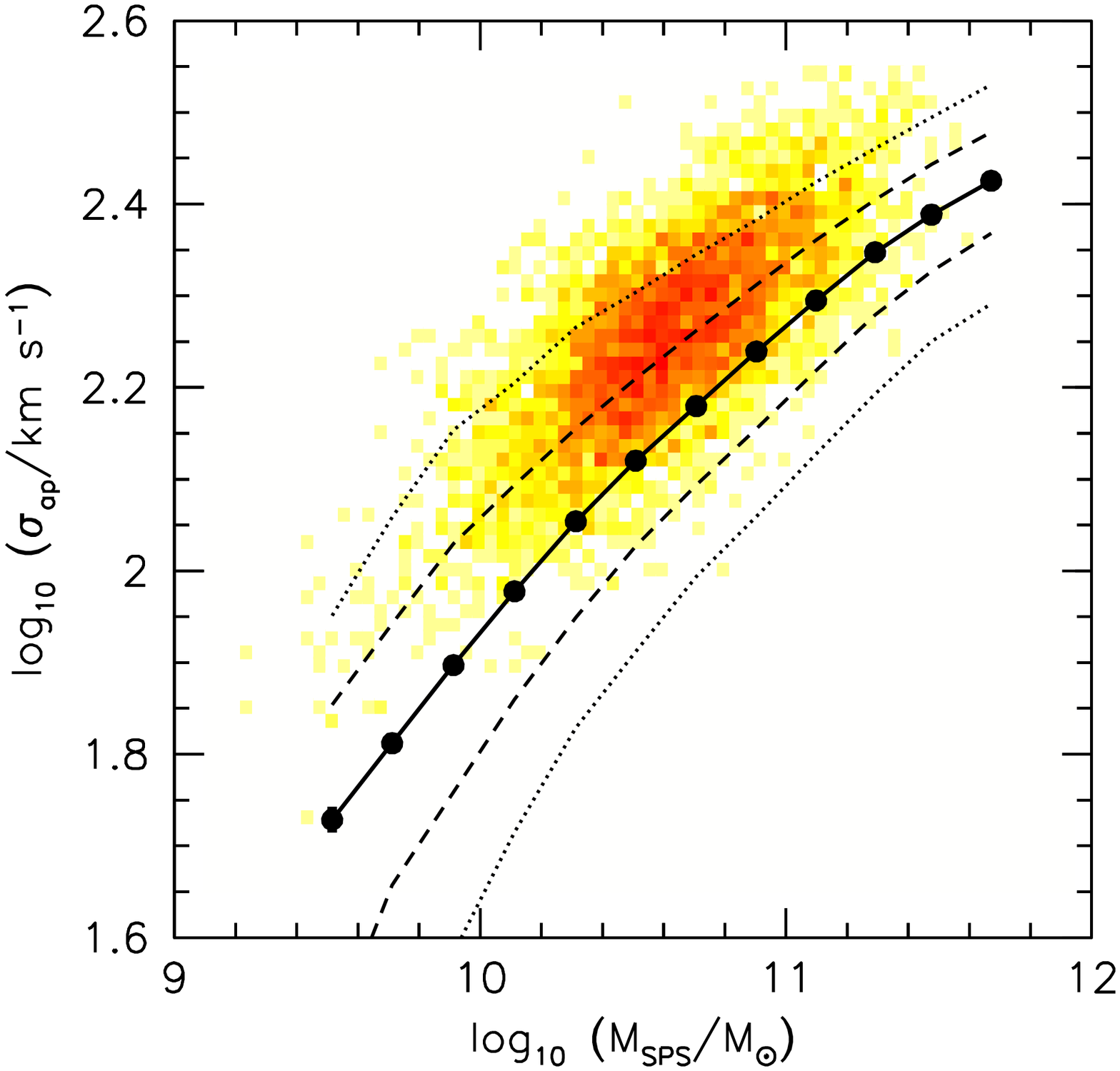,width=0.35\textwidth}
}
\caption{Size vs stellar mass and velocity dispersion vs stellar mass
  relations for our full and high-density samples of early-type
  galaxies. The relations for the full sample are shown with lines:
  median (solid); 16th and 84th percentiles (dashed); 2.5th and 97.5th
  percentiles (dotted). The color shading shows the relations for our
  high-density galaxies. In the size-mass relation the density selection is
  given by the red line with slope 1/2.}
\label{fig:rm}
\end{figure}

The velocity-mass and size-mass scaling relations for our full and
dense samples of early-type galaxies are shown in
Fig.~\ref{fig:rm}. The median relations of the full samples are shown
with points and solid lines, while the high density galaxies are shown
with a color scale.  In the size-mass relation our selection
corresponds to a line of slope 1/2.  All of the dense galaxies are
offset by at least $1\sigma$ from the median size-mass relation. In
the velocity-mass plane the dense galaxies have a slope of 0.25 and
a are offset to high velocity dispersions.

%% FIGURE 1

%% FIGURE 2
\begin{figure}
\centerline{
\psfig{figure=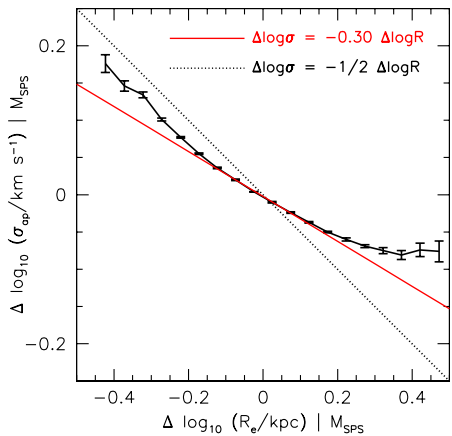,width=0.35\textwidth}
}
\caption{Correlation between residuals of the $\sigma_{\rm ap}-\Msps$
  and $R_{\rm e}-\Msps$ relations.  The virial fundamental plane has a
  slope of $-1/2$ (dotted line). The observations are given by the
  black line.  For $R_{\rm e}-\Msps$ residuals within $\pm0.2$ dex
  (which is 95\% of galaxies) the relation is linear with
  $\Delta\log\sigma_{\rm ap} = -0.30\, \Delta\log R_{\rm e}$,
  recovering the well known tilt of the fundamental plane. However,
  for larger size offsets the data clearly depart from a linear
  relation, representing curvature to the fundamental plane.}
\label{fig:dvr}
\end{figure}

%%%%%%%%%%%%%%%%%%%%%%%%%%%%%%%%%%%%%%%%%%%%%%%%%%%%%%%%%%%%%%%%%%%%%%
%% SECTION 3: Evidence for mass follows light
%%%%%%%%%%%%%%%%%%%%%%%%%%%%%%%%%%%%%%%%%%%%%%%%%%%%%%%%%%%%%%%%%%%%%%

\vspace{-0.6cm}
\section{EVIDENCE FOR MASS FOLLOWS LIGHT}
\label{sec:MFL}
The assumption of mass-follows-light (MFL) is sometimes used when
modeling the stellar kinematics of luminous early-type galaxies (e.g.,
Cappellari \etal 2006).  There are two issues with MFL models: 1) MFL
models only provide an upper limit to the stellar mass due to the
unknown dark matter fraction; 2) If there is a significant amount of
dark matter, then the assumption of MFL likely fails and the dynamical
masses will be biased. One can get around these two problems by
selecting galaxies for which the dark matter fraction is
negligible. To select such a sample we use the correlations between
velocity and size at fixed stellar mass.

The simplest mass follows light models have the scaling
$\sigma^2(R)\propto \Mstar/R$.  It is well known that the Fundamental
Plane is tilted with respect to this model. Here we express the FP
relation as the correlation between residuals from the velocity
dispersion vs stellar mass ($\Delta\log \sigma_{\rm ap}$) and
effective radius vs stellar mass ($\Delta\log R_{\rm e}$)
relations. We label the slope of the correlation as $\dvr\equiv
\Delta\log \sigma_{\rm ap}/\Delta\log R_{\rm e}$.  The value of $\dvr$
is related to the dark matter fraction within the radius the
velocities are measured (Courteau \& Rix 1999; Dutton \etal
2007). Models with no dark matter nominally have $\dvr=-1/2$, while
dark matter dominated galaxies have $\dvr \gta 0$.

In Fig.~\ref{fig:dvr} the solid black line shows the FP relation for
all early-type galaxies. Fitting a linear relation gives a slope $\dvr
\simeq -0.300\pm0.001$ (red line) which is shallower than the virial
prediction (dotted line). This recovers the well known tilt of the FP.
It is also apparent that the local value of $\dvr$ varies
monotonically with size offset. Galaxies with the most negative
offsets have $\dvr \simeq -0.5$, while galaxies with the most positive
offsets have $\dvr \simeq 0.0$. This trend is qualitatively consistent
with the expectations for galaxies embedded in extended dark matter
haloes.

\begin{figure}
\centerline{
\psfig{figure=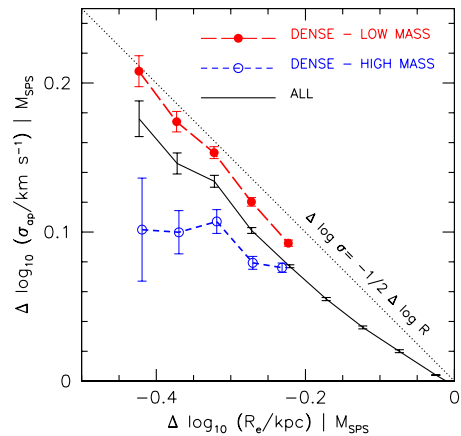,width=0.35\textwidth}
}
\caption{Correlation between residuals of the $\sigma_{\rm ap}-\Msps$
  and $R_{\rm e}-\Msps$ relations for dense galaxies.  For reference,
  the solid line shows the data for all galaxies from
  Fig.~\ref{fig:dvr}, but zoomed into the upper left quadrant
  ($\Delta\log \sigma_{\rm ap} > 0$, $\Delta\log R_{\rm e} < 0$). The
  red and blue points show the relation for dense galaxies
  ($\Sigma_{\rm e} > 2500 \,\Msun\,\pc^2$) split into high and low
  masses at $\Msps = 10^{10.8}\Msun$.  The low mass galaxies (red
  points) have a slope equal to the virial fundamental plane which
  suggests that they are baryon dominated within an effective radius.}
\label{fig:dvr2}
\end{figure}

From Fig.~\ref{fig:dvr} we see that galaxies with $\Delta\log R_{\rm
  e}\lta -0.2$ have a local slope of $\dvr \simeq -0.5$ and thus are
consistent with the assumption of MFL. This offset is the motivation
behind our particular density selection as shown in Fig.~\ref{fig:rm}.
The colored points in Fig.~\ref{fig:dvr2} show the FP relation for
galaxies in our dense sample, split into high and low stellar masses.
The low mass galaxies in our sample, $9.6 \lta \log_{10}(\Mstar/\Msun)
\le 10.8$, (red points) have $\dvr \simeq -0.5$ which suggests that
these galaxies are dominated by stars within an effective radius
(Courteau \& Rix 1999). However, the massive galaxies, $10.8 <
\log_{10}(\Mstar/\Msun) \lta 11.6$, (blue points) have a weaker
correlation, suggesting that mass does not follow light.

The possibility that dark matter follows the light in the inner
regions of elliptical galaxies has been discussed by Thomas \etal
(2011). However, this scenario seems unlikely for our dense galaxies,
since at the effective radius the slopes of the stellar mass profiles
are ${\rm d}\ln\rho_{\rm star}/{\rm d}\ln r\sim -2.5$, while the dark
matter is expected to have ${\rm d}\ln \rho_{\rm DM}/{\rm d}\ln r
>-2$.

%% FIGURE 3

Thus in what follows we can only strictly justify MFL models for the
lower mass dense galaxies.  However, using relations from Macci\`o
\etal (2008) and Dutton \etal (2010), we estimate the dark matter
fractions within $\Re$ to be just $\simeq 5\%$.  
We will still compute MFL models for the high mass dense galaxies,
with the caveat that dark matter may provide a non-negligible
contribution to the dynamical masses.  Our dense-low mass sample
consists of $\simeq 4000$ galaxies with median SPS masses $\Msps
\simeq 3\times 10^{10}\Msun$, velocity dispersions $\sigma_{\rm
  ap}=168^{+37}_{-34}\kms$, and fiber aperture radii of $R_{\rm
  ap}\simeq 1.5 R_{\rm e}$. The relatively large aperture radii
results in an additional benefit when deriving dynamical masses from
aperture velocity dispersions.  For small apertures relative to the
effective radius (e.g., one eighth of an effective radius), the
derived dynamical masses are strongly dependent on the anisotropy
profile. When the aperture is larger than the effective radius the
anisotropy is only of secondary importance.

%%%%%%%%%%%%%%%%%%%%%%%%%%%%%%%%%%%%%%%%%%%%%%%%%%%%%%%%%%%%%%%%%%%%%%
%% SECTION 4: Dynamical mass vs Stellar mass
%%%%%%%%%%%%%%%%%%%%%%%%%%%%%%%%%%%%%%%%%%%%%%%%%%%%%%%%%%%%%%%%%%%%%%

\vspace{-0.6cm}
\section{Dynamical mass vs Stellar Population Synthesis  Mass}
\label{sec:results}

For each galaxy in our high-density sample we calculate a dynamical
mass using the following procedure.  Given a galaxy light profile
(parameterized by circularized and deprojected S\'ersic $n=4$ and
$n=1$ components --- see Dutton \etal 2011a) with mass normalized to
$\Msps$, we calculate the radial velocity dispersion profile by
solving the spherical Jeans equations assuming a constant anisotropy,
$\beta$ (e.g., Binney \& Mamon 1982).  We then compute the projected
velocity dispersion (convolved with 1.4 arcsec FWHM seeing) within the
3 arcsec diameter aperture used by SDSS. Finally, we compute a
dynamical mass by scaling the model aperture velocity dispersion to
match the observed aperture velocity dispersion: $M_{\rm dyn}=M_{\rm
  SPS} (\sigma_{\rm ap,obs}/\sigma_{\rm ap,model})^2.$

Note that since the model velocity dispersions scale as $\Msps^{1/2}$
(because we assume mass-follows-light) our dynamical masses are
actually independent of the SPS masses.  Our procedure also takes into
account structural non-homology, which can significantly effect
dynamical masses (e.g., Taylor \etal 2010).
While our models do not account for rotation, Cappellari \etal (2006)
have shown that two and three integral models yield dynamical masses
that are consistent with those obtained from a simple virial relation
$\Mdyn \propto \sigma^2_e R_e$ for both fast- and slow-rotators, which
thus supports our use of spherical Jeans models. In addition we have
verified that the dynamical masses show no correlation with axis
ratio.

For isotropic orbits ($\beta=0$) the median offset between dynamical
and SPS masses for the full dense sample is $\Delta_{\rm IMF}\equiv
\log(\Mdyn/\Msps)=0.234\pm0.003$. For the low- and high-mass
subsamples we find $\Delta_{\rm IMF}=0.233\pm0.003$, and $\Delta_{\rm
  IMF}=0.236\pm0.005$, respectively.  As expected the dynamical masses
are only weakly dependent on the assumed velocity anisotropy
($\beta$). For reasonable values of velocity anisotropy ($0.0 \lta
\beta \lta 0.5$) for early-type galaxies (Gerhard \etal 2001; Koopmans
\etal 2009) the dynamical masses differ by just 4\% (with lower
dynamical masses for higher $\beta$). Thus, on average, our results
are in excellent agreement with a Salpeter IMF
($\DeltaIMF\simeq0.23$), but are strongly inconsistent with lighter
Milky-Way IMFs (e.g., Kroupa 2001; Chabrier 2003), or heavier IMFs
found in the centers of more massive ellipticals (van Dokkum \& Conroy
2010).

%% FIGURE 4
\begin{figure}
\centerline{
\psfig{figure=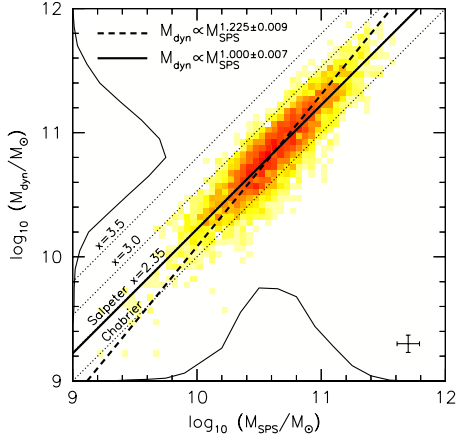,width=0.35\textwidth}
}
\caption{Relation between dynamical mass, $M_{\rm dyn}$, assuming
  $\beta=0$, and SPS stellar mass, $M_{\rm SPS}$, assuming a Chabrier
  (2003) IMF, for dense galaxies. 
  The relation between dynamical and SPS masses has a slope of
  $1.000\pm 0.007$ (for a forward fit, solid line), and
  $1.225\pm0.009$ (for an inverse fit, dashed line). The zero point
  offset is in excellent agreement with a Salpeter IMF.  The dynamical
  masses are, on average, inconsistent with lighter Milky-Way type
  IMFs (Chabrier), or bottom heavy ($x=3.0$, $x=3.5$) IMFs.}
\label{fig:mdyn-mstar}
\end{figure}

\vspace{-0.4cm}
\subsection{Variation of the IMF with galaxy mass}
Fig.~\ref{fig:mdyn-mstar} shows the relation between dynamical and SPS
masses for $\beta=0$. The dotted lines show the relations expected for
various IMFs, ranging from Chabrier to $x=3.5$. Due to the non-uniform
distribution of $\Msps$ (which is well approximated by a log-normal),
and the scatter in $\Mdyn/\Msps$ of $\simeq 0.15$ dex, the slope of
the true relation between $\Mdyn$ and $\Msps$ will depend on the
origin of the scatter.  If the scatter is entirely due to $\Mdyn$ the
slope is $1.000\pm0.007$ (solid line), while if the scatter is
entirely due to $\Msps$ the slope is $1.225\pm0.009$ (dashed line). In
practice there are likely errors on both masses, and the true slope of
the relation between $\Mdyn$ and $\Msps$ will fall between these two
limiting cases. The nominal measurement errors on $\Msps$ of $0.09$
dex, imply a true relation of $\Mdyn\propto \Msps^{1.1}$.

Since dynamical mass is proportional to the square of velocity
dispersion, our results are consistent with the notion that the IMF
depends on the velocity dispersion of the galaxy, with heavier IMFs in
higher dispersion galaxies (Treu \etal 2010; van Dokkum \& Conroy
2011).
Such a trend is also consistent with the lighter IMFs favored for
spiral galaxies (e.g., Bershady \etal 2011), which tend to have lower
velocities than ellipticals. However, since velocity dispersion
correlates with other physical galaxy properties (e.g., Graves \etal
2009), variation in the IMF may be driven by other properties of a
stellar population, such as age or metallicity.

%% FIGURE 5
\begin{figure}
\centerline{
\psfig{figure=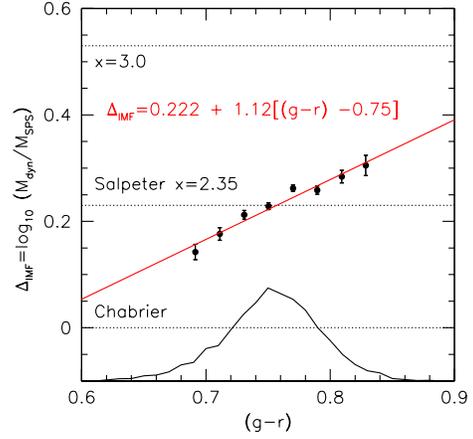,width=0.35\textwidth}
}
\caption{Dependence of the IMF offset parameter, $\Delta_{\rm
    IMF}\equiv \log_{10}(\Mdyn/\Msps)$, on $(g-r)$ galaxy color for
  low-mass dense early-types. Redder galaxies have heavier IMFs.}
\label{fig:delta-color}
\end{figure}

\vspace{-0.45cm}
\subsection{Variation of the IMF with galaxy color}
Fig.~\ref{fig:delta-color} shows that the IMF offset parameter is
correlated with color, with the sense that redder galaxies have
heavier IMFs.  This relation is given by
\begin{equation}
\label{eq:IMF-color}
\Delta_{\rm IMF}=0.222(\pm0.004) +1.12(\pm0.09)[(g-r)-0.75].
\end{equation}
Here the colors and luminosities have been k-corrected to $z=0$ (but
not evolution corrected).  Extrapolating Eq.~\ref{eq:IMF-color} to
bluer colors, a Chabrier IMF (i.e., $\Delta_{\rm IMF}=0$) occurs at a
color of $(g-r)=0.56$, which is typical for a $\Msps\sim3\times
10^{10}\Msun$ late-type galaxy (e.g., Fig.1 of Dutton \etal 2011a).

A color dependence to the IMF could plausibly be related to the idea
that the characteristic mass scale of the IMF is set by the Jeans mass
of the ISM, which depends on both temperature and density (Bate \&
Bonnel 2005; Larson 2005).  A density dependence would take the form
of heavier IMFs for stars formed in denser regions. Since more massive
galaxies formed their stars at higher densities, and more massive
galaxies have older and redder stellar populations, it follows that
galaxies that formed their stars at higher densities (with heavier
IMFs) should have redder colors.

%% FIGURE 6
\begin{figure}
\centerline{
\psfig{figure=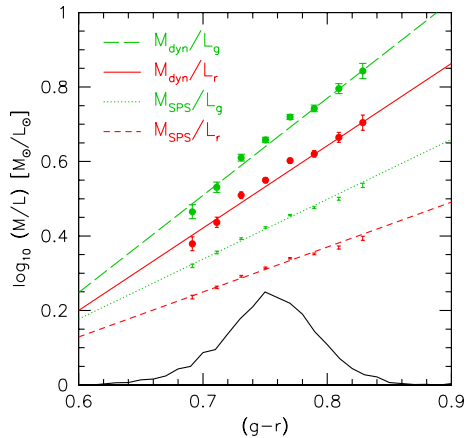,width=0.35\textwidth}
}
\caption{Relations between mass-to-light ratios and $(g-r)$ color for
  a model with isotropic orbits ($\beta=0$).  Dynamical mass-to-light
  ratios ($M_{\rm dyn}/L$) have stronger scalings with color than SPS
  mass-to-light ratios ($M_{\rm SPS}/L$). }
\label{fig:mdyn-l}
\end{figure}

\vspace{-0.4cm}
\subsection{Dynamical mass-to-light ratio vs color}
With the goal of providing a useful estimator of stellar mass-to-light
ratios, Fig.~\ref{fig:mdyn-l} shows the relations between dynamical
(and SPS) mass-to-light ratio vs $(g-r)$ color for a model with
isotropic orbits $(\beta=0)$. SPS mass-to-light ratios scale as
$\Msps/L_g \propto (g-r)^{1.61}$ and $\Msps/L_r \propto
(g-r)^{1.21}$. These slopes are very similar to those of Bell \etal
(2003), who report $1.52$ and $1.10$ for the $g$- and $r$-bands,
respectively.
The relations between dynamical mass-to-light ratios and color in the
$g$- and $r$-bands from Fig.~\ref{fig:mdyn-l} are given by the
following equations:
\begin{equation}
M_{\rm dyn}/L_{g} = 0.640(\pm0.003) +2.61(\pm 0.09)[(g-r)-0.75],
\end{equation} 
\begin{equation}
M_{\rm dyn}/L_{r} = 0.532(\pm0.003) +2.21(\pm 0.09)[(g-r)-0.75].
\end{equation} 
The steeper slopes of the $\Mdyn/L$ vs. $(g-r)$ compared to the
$\Msps/L$ vs $(g-r)$ relations confirm that the trend between IMF
offset and galaxy color shown in Fig.~\ref{fig:delta-color} is not due
to %potential
correlations between $\Msps$ and $(g-r)$.

%%%%%%%%%%%%%%%%%%%%%%%%%%%%%%%%%%%%%%%%%%%%%%%%%%%%%%%%%%%%%%%%%%%%%%
%% SECTION 5: SUMMARY
%%%%%%%%%%%%%%%%%%%%%%%%%%%%%%%%%%%%%%%%%%%%%%%%%%%%%%%%%%%%%%%%%%%%%%

\vspace{-0.6cm}
\section{Summary}
\label{sec:sum}

We have determined an absolute calibration of stellar mass-to-light
ratios from the densest $\simeq 3\%$ of early-type galaxies in the
local universe (redshifts $z\simeq 0.08$) from the SDSS DR7. The
fundamental plane scaling relations of these galaxies (with SPS masses
$\log_{10}(\Msps/\Msun)<10.8$, assuming a Chabrier IMF) are consistent
with the virial plane: $\sigmaap^2\propto \Msps/R_{\rm e}$, suggesting
that mass-follows-light. With the additional assumption that any dark
matter will not follow the light, we infer that these galaxies are
dominated by stars, i.e., they contain little or no dark matter within
an effective radius.

We compute dynamical masses using spherical Jeans models with constant
velocity anisotropy, $\beta$.  For isotropic orbits ($\beta=0$) the
dynamical masses are on average $0.233\pm0.003$ dex heavier than the
SPS masses obtained with a Chabrier (2003) IMF. For radially
anisotropic orbits with $\beta=0.5$ the median offset is just 0.02 dex
lower.  The dependence on the anisotropy is small because the velocity
dispersions are measured within a fairly large aperture: $\simeq 1.5
R_{\rm e}$.

The relation between dynamical and SPS masses,
$\Mdyn\propto\Msps^{\alpha}$, has a slope $1.00 \lta \alpha \lta 1.23$
depending on the source of the scatter in this relation, with a most
likely value of $\alpha=1.1$. Our results are thus consistent with,
but do not require, a dependence of the IMF on dynamical mass or
velocity dispersion.
We find a strong dependence between $\Mdyn/\Msps$ and $(g-r)$
color, with the sense that redder galaxies have heavier IMFs.  If this
trend reflects a more fundamental dependence of the IMF on the age and/or
metallicity of a stellar population, it implies that the IMF varies
within as well as between galaxies.

{\it Future applications of this method.}  This method has the
potential to be applied to galaxies at higher redshifts, and thus
probe the evolution of the IMF. Since the sizes of early-type galaxies
are known to decrease as we look backward in time (e.g., Trujillo
\etal 2006), we expect our method to be applicable to a larger
fraction of early-type galaxies, and specifically higher mass
galaxies, at higher redshifts.

\vspace{-0.5cm}
\section*{Acknowledgments} 
We thank Tommaso Treu, Charlie Conroy and Frank van den Bosch for
valuable discussions. A.A.D. acknowledges financial support from the
Canadian Institute for Theoretical Astrophysics (CITA) National
Fellows program.

%%%%%%%%%%%%%%%%%%%%%%%%%%%%%%%%%%%%%%%%%%%%%%%%%%%%%%%%%%%%%%%%%%%%%%
%%  REFERENCES
%%%%%%%%%%%%%%%%%%%%%%%%%%%%%%%%%%%%%%%%%%%%%%%%%%%%%%%%%%%%%%%%%%%%%% 

\label{lastpage}

\vspace{-0.5cm}
\begin{thebibliography}{}

%The Seventh Data Release of the Sloan Digital Sky Survey
\bibitem[Abazajian et al.(2009)]{2009ApJS..182..543A} Abazajian,
  K.~N., et al.\ 2009, \apjs, 182, 543

%Dark Matter Contraction and the Stellar Content of Massive Early-type
% Galaxies: Disfavoring "Light" Initial Mass Functions
\bibitem[Auger et al.(2010)]{2010ApJ...721L.163A} Auger, M.~W., Treu,
  T., Gavazzi, R., Bolton, A.~S., Koopmans, L.~V.~E., \& Marshall,
  P.~J.\ 2010, \apjl, 721, L163

  % The origin of the initial mass function and its dependence on the
  % mean Jeans mass in molecular clouds
\bibitem[Bate \& Bonnell(2005)]{2005MNRAS.356.1201B} Bate, M.~R., \&
  Bonnell, I.~A.\ 2005, \mnras, 356, 1201

%%Stellar Mass-to-Light Ratios and the Tully-Fisher Relation
\bibitem[Bell \&  de Jong(2001)]{2001ApJ...550..212B} Bell,  E.~F., \&
  de Jong, R.~S.\ 2001, \apj, 550, 212

  % The Optical and Near-Infrared Properties of
  % Galaxies. I. Luminosity and Stellar Mass Functions
\bibitem[Bell et al.(2003)]{2003ApJS..149..289B} Bell, E.~F., McIntosh, 
D.~H., Katz, N., \& Weinberg, M.~D.\ 2003, \apjs, 149, 289 

%Galaxy Disks are Submaximal
\bibitem[Bershady et al.(2011)]{2011ApJ...739L..47B} Bershady, M.~A., 
Martinsson, T.~P.~K., Verheijen, M.~A.~W., et al.\ 2011, \apjl, 739, L47 

% M/L and velocity anisotropy from observations of spherical galaxies,
% or must M87 have a massive black hole
\bibitem[Binney \& Mamon(1982)]{1982MNRAS.200..361B} Binney, J., \&
  Mamon, G.~A.\ 1982, \mnras, 200, 361

% The SWELLS survey. III. Disfavouring "heavy" initial mass functions for spiral lens galaxies
\bibitem[Brewer et al.(2012)]{2012arXiv1201.1677B} Brewer, B.~J.,
  Dutton, A.~A., Treu, T., et al.\ 2012, arXiv:1201.1677

% The SAURON project - IV. The mass-to-light ratio, the virial mass
% estimator and the Fundamental Plane of elliptical and lenticular
% galaxies
\bibitem[Cappellari et al.(2006)]{2006MNRAS.366.1126C} Cappellari, M., et 
al.\ 2006, \mnras, 366, 1126 

%Galactic Stellar and Substellar Initial Mass Function
\bibitem[Chabrier(2003)]{2003PASP..115..763C} Chabrier, G.\ 2003, PASP, 
115, 763 

%Maximal Disks and the Tully-Fisher Relation
\bibitem[Courteau \& Rix(1999)]{1999ApJ...513..561C} Courteau, S., \&
  Rix, H.-W.\ 1999, \apj, 513, 561

% Fundamental properties of elliptical galaxies
\bibitem[Djorgovski \& Davis(1987)]{1987ApJ...313...59D} Djorgovski,
  S., \& Davis, M.\ 1987, \apj, 313, 59

% Spectroscopy and photometry of elliptical galaxies. I - A new
% distance estimator
\bibitem[Dressler et al.(1987)]{1987ApJ...313...42D} Dressler, A.,
  Lynden-Bell, D., Burstein, D., et al.\ 1987, \apj, 313, 42

  % A Revised Model for the Formation of Disk Galaxies: Low Spin and
  % Dark Halo Expansion
\bibitem[Dutton et al.(2007)]{2007ApJ...654...27D} Dutton, A.~A., van
  den Bosch, F.~C., Dekel, A., \& Courteau, S.\ 2007, \apj, 654, 27

%The kinematic connection between galaxies and dark matter haloes
\bibitem[Dutton et al.(2010)]{2010MNRAS.407....2D} Dutton, A.~A.,
  Conroy, C., van den Bosch, F.~C., Prada, F., \& More, S.\ 2010,
  \mnras, 407, 2

% Dark Halo Contraction and the stellar initial mass function of early
% and late type galaxies
\bibitem[Dutton et al.(2011a)]{2011MNRAS.416..322D} Dutton, A.~A., Conroy, 
C., van den Bosch, F.~C., et al.\ 2011a, \mnras, 416, 322 

%	The SWELLS survey. II. Breaking the disk-halo degeneracy in
% the spiral galaxy gravitational lens SDSS J2141-0001
\bibitem[Dutton et al.(2011b)]{2011arXiv1101.1622D} Dutton, A.~A., Brewer, 
B.~J., Marshall, P.~J., et al.\ 2011b, \mnras, 417, 1621

  % Dynamical Family Properties and Dark Halo Scaling Relations of
  % Giant Elliptical Galaxies
\bibitem[Gerhard et al.(2001)]{2001AJ....121.1936G} Gerhard, O., 
Kronawitter, A., Saglia, R.~P., \& Bender, R.\ 2001, \aj, 121, 1936 

%	Dissecting the Red Sequence. I. Star-Formation Histories of
% Quiescent Galaxies: The Color-Magnitude versus the Color-sigma Relation
\bibitem[Graves et al.(2009)]{2009ApJ...693..486G} Graves, G.~J., Faber, 
S.~M., \& Schiavon, R.~P.\ 2009, \apj, 693, 486 

% The Structure and Dynamics of Massive Early-Type Galaxies: On
% Homology, Isothermality, and Isotropy Inside One Effective Radius
\bibitem[Koopmans et al.(2009)]{2009ApJ...703L..51K} Koopmans, L.~V.~E., 
Bolton, A., Treu, T., et al.\ 2009, \apjl, 703, L51 

%On the variation of the initial mass function
\bibitem[Kroupa(2001)]{2001MNRAS.322..231K} Kroupa, P.\ 2001, \mnras,
  322, 231

%Thermal physics, cloud geometry and the stellar initial mass function
\bibitem[Larson(2005)]{2005MNRAS.359..211L} Larson, R.~B.\ 2005, \mnras, 
359, 211 

% Concentration, spin and shape of dark matter haloes as a function of
% the cosmological model: WMAP1, WMAP3 and WMAP5 results
\bibitem[Macci{\`o} et al.(2008)]{2008MNRAS.391.1940M} Macci{\`o},
  A.~V., Dutton, A.~A., \& van den Bosch, F.~C.\ 2008, \mnras, 391,
  1940

\bibitem[Salpeter(1955)]{1955ApJ...121..161}  Salpeter,  E.~E.\  1955,
\apj, 121, 161

% A Catalog of Bulge+disk Decompositions and Updated Photometry for
% 1.12 Million Galaxies in the Sloan Digital Sky Survey
\bibitem[Simard et al.(2011)]{2011ApJS..196...11S} Simard, L., Mendel,
  J.~T., Patton, D.~R., Ellison, S.~L., \& McConnachie, A.~W.\ 2011,
  \apjs, 196, 11

%On the Masses of Galaxies in the Local Universe
\bibitem[Taylor et al.(2010)]{2010ApJ...722....1T} Taylor, E.~N.,
  Franx, M., Brinchmann, J., van der Wel, A., \& van Dokkum, P.~G.\
  2010, \apj, 722, 1

  %% Dynamical masses of early-type galaxies: a comparison to lensing
  %% results and implications for the stellar initial mass function
  %% and the distribution of dark matter
\bibitem[Thomas et al.(2011)]{2011MNRAS.415..545T} Thomas, J., Saglia,
  R.~P., Bender, R., et al.\ 2011, \mnras, 415, 545

%The Initial Mass Function of Early-Type Galaxies
\bibitem[Treu et al.(2010)]{2010ApJ...709.1195T} Treu, T., Auger,
  M.~W., Koopmans, L.~V.~E., Gavazzi, R., Marshall, P.~J., \& Bolton,
  A.~S.\ 2010, \apj, 709, 1195

%The Size Evolution of Galaxies since z~3: Combining SDSS, GEMS, and FIRES	
\bibitem[Trujillo et al.(2006)]{2006ApJ...650...18T} Trujillo, I., 
F{\"o}rster Schreiber, N.~M., Rudnick, G., et al.\ 2006, \apj, 650, 18 

  %	A substantial population of low-mass stars in luminous
  % elliptical galaxies
\bibitem[van Dokkum \& Conroy(2010)]{2010Natur.468..940V} van Dokkum,
  P.~G., \& Conroy, C.\ 2010, Nature, 468, 940

  %	Confirmation of Enhanced Dwarf-sensitive Absorption Features
  % in the Spectra of Massive Elliptical Galaxies: Further Evidence
  % for a Non-universal Initial Mass Function
\bibitem[van Dokkum \& Conroy(2011)]{2011ApJ...735L..13V} van Dokkum,
  P.~G., \& Conroy, C.\ 2011, \apjl, 735, L13

%The Sloan Digital Sky Survey: Technical Summary
\bibitem[York et al.(2000)]{2000AJ....120.1579Y} York, D.~G., et al.\
  2000, \aj, 120, 1579


\end{thebibliography}
\end{document}